# Entropy NAND: Early Functional Completeness in Entropy Networks

## Forrest Fabian Jesse[1,2*]

[1] School of Computer and Information Technology, Beijing Jiao Tong University, Beijing, 100044, China

[2] Bio-X Institutes, Shanghai Jiao Tong University, Shanghai, 200240, China

* Author to whom correspondence should be addressed; email: jesse@jesse.org;

Abstract: An observer increases in relative entropy as it receives information from what it is observing. In a system of only an observer and the observed, an increase in the relative entropy of the observer is a decrease in the relative entropy of the observed. Linking together these directional entropy disequilibriums we show that NAND and NOR functionality arise in such networks at very low levels of complexity.

## 1. Introduction

We describe a simple structure of directed entropy transfer which we label an observation. This observation is the simplest network pattern of directional entropy transfer consisting of an observer and an observed. The characterization of observation as a thermodynamic transfer has been well described previously [1, 2]. The resulting entropy generation from observation can transfer [3] in a network [4-8]. In the tiny networks we discuss, the resulting entropy differentials can be used to construct or find a NAND or a NOR gate. NAND and NOR can both be found and are present in the same network structure, distinguished only by changing the threshold of the output. We focus only on the NAND gate here. The thermodynamically actuated NAND gate is trivial to implement which points to the possibility that it exists in nature.

Observation is a source of entropy increase if it requires two states for an observer (a previous state and a changed state) as the observer acquires information or is surprised. The unit surprisal [9] can be used to measure how much information is acquired, where complete lack of surprise is a complete lack of information transfer. The smallest surprise is a nat, the unit of entropy in base $e$. Entropy creation from observation has been described to have a minimum limit [1,10,11,12] of $k_B$ T ln 2 ($k_B$ is the Boltzmann constant, T is absolute temperature, 2 represents the before and the after state) and this also defines the minimum energy required to change a bit of information. The minimum limit has since been experimentally shown [12] to be valid. This entropy change has been characterized in the Brillouin description of *negentropy* [1, (eq. 3)] for an observer which remains isothermal by radiating the entropy increase as thermal energy. The negentropy principal builds on previous work (Szilard [2] and others) stating that the entropy of the observer is negative as entropy is transferred to the observer's environment.

Instances of the generation and transfer of entropy through observation can be linked together to produce an observation network model which we illustrate in *figure 1*. Entropy network structures have been investigated previously [8,13,14]. A description of entropy transfer can be found in the Schreiber *transfer entropy* [3] which has been thermodynamically characterized by Prokopenko, Lizier & Price [15]. We describe a discrete entropy transfer between two idealized elements which are entropy containers.

Networks constructed from observations can be functionally complete, which we show in this article by constructing a NAND gate from entropy transfers.

## 2. Results and Discussion

*2.0. Observers*

We propose our version of an observer as a discrete element which can change. By discrete we mean having no relation with any other element, except through the relationship type which we describe as an observation. We define element to be that which generates 1 nat of entropy in the observer when observed. The element is the prototype for possible thermodynamic disequilibrium, initially without internal differentiation. The interior of our prototypical element is homogeneous, having no features to compare and so having no internal information, and no entropy. The first differentiation in an element is the instantiation of disequilibrium that is a directed transfer of entropy. As stated previously, entropy creation from observation has been described to have a minimum limit [1,10,11,12] of $k_B$ T ln 2. We can set an appropriate temperature so that $k_B = 1$ which simplifies the description of the entropy of the observer to:

$$T \ln m \tag{1}$$

where *m* is the quantity of states of the observer. Before observation, *m* = 1, through observation *m* = 2 at least momentarily until the entropy is transferred into the environment. We can represent this element as a pattern that has no features, illustrated in *figure 1.a* as a circle where *m* = 1; and as a divided circle where *m*=2. Our small model system is shown in *figure 1* and *2*. We illustrate the element as a pattern that has no features in *figure 1.a* as a circle. An element in internal entropy disequilibrium as represented as a circle split into parts, *figure 1.b* labeled *b1, b2*, illustrated with lines drawn through the circle enumerating the component states of disequilibrium. An example shows two states and the simplest entropy disequilibrium in *figure 1.b*.

2.1. Observation of an object

Our version of observation of an object is a process in which the observer gains at least one bit of information, and which is an entropy change. In a system of only the observer and the observed, this relative entropy change is described by 1 nat. We can find a physical description of the relation (measuring temperature T, S is total summed entropy) as S=E/T as stated by Brillouin [1, (eq. 40)] where the total entropy increase from a single observation is described.

We equate the smallest entropy increase to the simplest differentiation of a discrete element, resulting in an entropy increase of 1 nat. The entropy increase we describe as a split of the observer into two states which we can describe as "before observation" and "after observation".

In the course of observation the complexity of the element is momentarily increased, which increases its entropy in relation to those observers which observe it. This entropy must be dissipated to the environment if the element is to remain isothermal, we represent the environment as another discrete element as defined previously, this is illustrated in the transfer network in *figure 1.c*

The graph of binary entropy H(p) shows entropy peaks between the beginning and ending states in a system of one bit expressing a state change. This graph can be verified experimentally on a statistical basis. In this article we describe a "before" and an "after" state, and propose that the transition between these states follows the binary entropy graph through time, yielding maximum entropy at the theoretical midpoint of the transition. The transition between states for our simple element model is a momentary peak at maximum entropy, which we describe as a Kronecker delta function that is 1 only at 0 (0 being the theoretical midpoint of the transition). For illustrative purposes we also describe in section 3 a physical example using ice cubes as entropy containers, in which the transition would need to occur over a longer span of time.

2.2. *Entropy transfer*

The minimum entropy transfer on observation is 1 nat which is a relative entropy reduction for the observed:

$$\text{Minimum Entropy Transfer} = 1 \text{ nat} \qquad (2)$$
$$\text{Minimum Energy Required} = k_B \text{ Temperature ln 2 (Joules)}$$

At stable temperatures (isothermal conditions, in Joules) [10] this defines the minimum energy required to transfer one bit of information. For an isothermal system, if entropy has increased, such entropy must be absorbed by the environment, otherwise the system is not isothermal. When we model this transfer with the Kronecker delta, the absorption of new entropy appears to occur instantly for elements which are connected directly (with a network distance of 1)

*2.3. Entropy of erasure*

If the system in which a bit is changed is small enough, and it has no memory, then changing a bit is deleting information. This deletion of information by the Landauer principle will require the dissipation of heat of at least $k_B T \ln 2$ of heat per lost bit [13]. This dissipation interacts with the environment of the observer. In a small system where the environment is composed of only a single discrete element (*figure 1.c*), the dissipated entropy must be absorbed by this single element.

*2.4. Shifting entropy from the classified to the classifier*

When a collection is classified, entropy shifts from the classified to the classifier, this process of classification from Kullback and Leibler [16] is an entropy reduction in the classified, and an entropy increase in the classifier. A single discrete element, which has capacity for only holding one bit of information observes another element: if this discrete element is to receive information from what it has observed, it must change to another state which has a bit of information correlated with what it has observed. This correlation with what it has observed is a relationship which in the simplest case is unidirectional. The smallest element is just 1 bit, with no memory. A memoryless 1 bit element may have no record of the state change, and so by the Landauer principle we will find an increase in entropy in the environment which is the classification system. This is again *figure 1.c*.

*2.5. Observation networks map entropy transfers*

We can conceptualize observation networks to map entropy transfers where at least one transfer involves an observer. Entropy is a convenient quantity to map as it transitions through many networks which are traditionally separate.

We can build observation networks as graph-like descriptions of the pathways of change between elements, where the elements can then model objects, people, proteins, neurons or other graphable systems. Graphing tools for visualizing and using biological interaction networks [4-7], have been developed across the span of scales: from molecular and protein level interaction models [6] to models of human interaction in social orders. Biological architectures in general have been abstractly characterized as network models [7,8] .

In the following diagrams we describe the creation of an element's relationships with other elements. These elements are an accounting device which we can use to trace entropy transfer. An element can split into two elements (before and after) in the creation of a nat of entropy, which is an observation and the simplest relationship.

Entropy before observation is Temperature ln 1 =0, as there are no relationships in the system (*figure 1.a*). We illustrate entropy networks elaborating possible networks up to network distance d = 2 (*figures 1.c* and *1.d*; *figure 2.d* show to d = 2 ). Two simultaneous observations (d = 1) are sufficient to create an entropy NAND gate (*figure 3*). Deeper exploration of observation network sequences is interesting but remains unexplored here.

**Figure 1. Entropy Circuit**. From top to bottom is shown a description of observation as targeted entropy generation and transfer. The circle represents an element with a single state. The divided circle represents an element with multiple states (before observation and after observation), being divided by a line for each additional state. The arrow indicates the target of observation, the element at the origin of the arrow experiences a split into before and after states on observation represented by the arrow crossing the circle to show it divided.

|   | symbol | example | description |   |
|---|---|---|---|---|
| a) |   | $a$  $b$ | Two elements. | $a = T \ln 1 = 0$ nat<br>$b = T \ln 1 = 0$ nat |
| b) | observation | $a$  $b_1$ / $b_2$ | Element $b$ observes element $a$, causing $b$ to split into 2 states $\{b_1, b_2\}$. | $a = T \ln 1 = 0$ nat<br>$b = T \ln 2 = 1$ nat |
| c) | observation train | $a$  $b_1$ / $b_2$  $\{env_1, env_2\}$ | Element $b$ is observed by its environment, dissipating its entropy as the two states $\{env_1, env_2\}$, to be measured as an absolute temperature change. | $a = T \ln 1$<br>$b = T \ln 2$<br>$env \geq T \ln 2$ |

*2.6. Entropy transfer classes*

We can classify entropy network structures according to their entropy transfer directionality. This can be a simple means to classify simple patterns through entropy networks.

Entropy networks are diverse. Entropy-driven organization occurs at least at the molecular scale [17]. Entropy is directed by people, ubiquitous in the practice of thermodynamic industry. People are sensitive to entropy at large scales [18] and directly through sensory perception in the course of pattern recognition. These areas all express entropy networks, while a common classification of entropy network components throughout these systems is lacking. Classification by total entropy [19], has been examined [20] to search for information. But using a measure of total entropy tells us nothing about what the net connects to and how it is formed. Classification by connectivity has been carried out historically in the practice of geometry and related studies. Unsurprisingly, there are several basic transfer patterns from which other patterns can be built. We classify them by how they transfer entropy, by linking together instances of observation.

We describe two classes, which we label first order and second order, the order is an indication of the quantity of observations that compose the network. We stop at two. The first order contains only one pattern of observation, the single observation. The second order contains four patterns, which are all possible patterns for two observations. In the second order, the only difference between networks is the directionality of entropy transfer.

*2.7. A first order classification (all possible patterns for one observation)*

1) The single observation. This is the simplest discrete entropy transfer describing a single element splitting into two states on observation of another element, creating entropy disequilibrium of 1 nat.

*2.8. A second order classification (all possible patterns for two observations)*

This second order of classification is all possible patterns for two observations. The patterns have equal complexity but each transfers entropy differently.

1) Sequential entropy transfer (*figure 2.d*)
2) Loop of entropy transfer (*figure 2.a*)
3) A pattern in which two elements observe a common element, we label it $e^>$. (*figure 2.b*)
4) A pattern in which an element is split through observation to observe two other elements, we label it $s^<$. (*figure 2.c*)

**Figure 2. Two Observation Entropy Transfer Patterns**. From top to bottom, we show all possible patterns for two observations, which produces four distinct patterns each of which has a distinct entropy transfer direction.

| | symbol | example | description | |
|---|---|---|---|---|
| a) | loop | $a_1$ $b_1$ / $a_2$ $b_2$ | Element $b$ splits on observation of $a$; $a$ splits on observation of $b$. | |
| b) | e> | | Element $e$ is observed by two elements, decreasing relative entropy for $e$. | $e = T \ln 1$ <br> $a = T \ln 2$ <br> $b = T \ln 2$ |
| c) | s< | | Element $s$ in observation of two elements, increasing relative entropy for $s$. | $s = T \ln 4$ <br> $a = T \ln 1$ <br> $b = T \ln 1$ |
| d) | observation train | $p$ $b$ $c$ | Element $c$ observes $b$ which observes $p$ in a linear train of entropy transfer. | $p = T \ln 1$ <br> $b = T \ln 2$ <br> $c \geq T \ln 2$ |

*2.9. Early functional completeness*

At very low complexity, we find that the second order observation network patterns exhibit functional completeness. Excepting the loop, the second order (two observation) class is the functionally complete NAND gate. Each of the patterns in this class serves as one of the four possibilities of the gate.

NAND state (inputs:11; output:0 ) is shown by the s< pattern shown in *figure 2.c*, which has two elements being observed and so has lower relative entropy while their observer, being split through before and after states has higher entropy.

NAND state (inputs:00; output:1) is shown in the e$^>$ pattern (*figure 2.b*), in which the central e element is of low relative entropy because it is being observed by two other elements of higher relative entropy.

The next two NAND states are represented by the two other possibilities that exist for the sequential, train-like observation patterns that are *btrain* (inputs:01; output:1) and *atrain* (inputs:10; output:1). We can note that these last two patterns, being sequential, can only be distinguished (front to back or back to front) if the NAND gate is embedded within another network, which is appropriate because a logic gate requires input and output to be a gate.

**Figure 3. NAND**. From top to bottom each pattern represents a logical NAND gate state. The output of the gate is read from *o*. The inputs are *a* and *b*. The outputs should be thresholded so that the logical false $\leq \tfrac{1}{2}$ and logical true $> \tfrac{1}{2}$.

| symbol | example | a | b | o | |
|---|---|---|---|---|---|
| e> | | 0 | 0 | 1 | $a = \widehat{o_1}$<br>$b = \widehat{o_2}$<br>$a = T \ln 2$<br>$b = T \ln 2$<br>$o = T \ln 1$ |
| b, o, a train | | 0 | 1 | 1 | $a = \hat{o}$<br>$o = \hat{b}$<br>$a \geq T \ln 2$<br>$b = T \ln 1$<br>$o = T \ln 2$ |
| a, o, b train | | 1 | 0 | 1 | $b = \hat{o}$<br>$o = \hat{a}$<br>$b \geq T \ln 2$<br>$a = T \ln 1$<br>$o = T \ln 2$ |
| s< | | 1 | 1 | 0 | $o = \hat{a}, \hat{b}$<br>$a = T \ln 1$<br>$b = T \ln 1$<br>$o = T \ln 4$ |

In the following expressions we use the caret notation (a hat) to represent that an element is the observation of another element, for example $a = \hat{o}_1$ means that $a$ is the observation of $o_1$.

$$e^> = \begin{cases} a = \hat{o}_1 \,; a_S \geq T \ln 2 \\ b = \hat{o}_2; b_S \geq T \ln 2 \\ o_S = T \ln 1 \end{cases} \quad (3)$$

$$s^< = \begin{cases} a_S = T \ln 1 \\ b_S = T \ln 1 \\ o = \hat{a}\,\hat{b}; \, o_S \geq T \ln 4 \end{cases} \quad (4)$$

$$btrain = \begin{cases} a = \hat{o}; a_S \geq T \ln 2 \\ b_S = T \ln 1 \\ o = \hat{b}; \, o_S \geq T \ln 2 \end{cases} \quad (5)$$

$$atrain = \begin{cases} a_S = T \ln 1 \\ b = \hat{o}; b_S \geq T \ln 2 \\ o = \hat{a}\,; o_S = T \ln 2 \end{cases} \quad (6)$$

## 3. Discussion

To physically embody this NAND system would require the embodiment of elements which can contain entropy. In our simple example we use discrete elements which have discrete entropy values, these values are controlled by the direction of entropy transfer between elements. A macro-scale implementation of this could be constructed, but would become statistical as the thermodynamic transfers would be of higher energies and not discrete observations.

In a possible experimental implementation our simple elements could be embodied by ice cubes which are thermodynamic heat reservoirs, and entropy transfer could be arranged by heat pumping, thermally conductive paths between the reservoirs. Two ice cubes are inputs which are connected to a central ice cube by a thermal pump, and the central ice cube is a NAND output. For our macro-scale implementation, the thermal coupling must be a directionally controllable entropy pump, which could be implemented as a heat pump (thermal transfer engine).

The three elements that compose our NAND gate are then three ice cubes. Entropy increase, considering constant volume and pressure would then be a temperature increase, and high entropy ice cubes would melt. Low entropy ice cubes would be solid. In the two-link (two observation) network that we have described, if we are to drain the entropy from two ice cubes (*figure 2.b*), the central ice cube (represented in figure 2.b as element e) must absorb the entropy, and melt. If we are to drain the entropy from the central ice cube into the other two (*figure 2.b*), they must increase in entropy and melt while the central ice cube (e) freezes.

We note that within a quantity of exactly 2 observations, NAND is possible with output thresholded so that the logical false ≤ ½ and logical true > ½. NOR is possible with a slight modification of this threshold. AND, OR, XAND and XOR are unreachable with two observation networks, they require more than 2 observations for implementation. That discrete observation entropy networks can attain functional completeness almost instantly may be but a curiosity. However, the statistical, thermodynamic NAND/NOR produces the same functionality in the macro scale, and the thermodynamic gate would share in the physical ubiquity of thermodynamic transfer.

The energy associated with a single observation and it's relation to relative entropy in quantum models was already noted by Brillouin in 1953. It has been increasingly accounted for in quantum collapse models [21-22]. Huaiyu Zhu [23] and others have described the entropy of photon-mass absorption, and though this description is a larger scale entropy transfer than that in the observation networks model presented here, photon-mass entropy transfer might allow an experimentally feasible micro-scale characterization of entropy transfer at the quantum scale. Entropy reduction is a collapse of multiple states to fewer states. This feature of collapse, in conjunction with entropy NAND perhaps points to the possibility of fine grained computational style structures with physical results.

The thermodynamic NAND which follows from entropy NAND points to the existence of scale invariant computationally functional entropy transfer networks and proposes ubiquitous computation occurring throughout such networks.

## 4. Conclusions

Thermodynamic entropy networks are ubiquitous. Our simple characterization of entropy networks shows that at a very early stage of complexity functional completeness can be produced. As noted in the introduction, the thermodynamic NAND gate is trivial to construct, and because of this simplicity it may be present in nature.

**Conflicts of Interest**

The authors declare no conflict of interest.